\documentclass[a4paper]{amsart}   

\usepackage[english]{babel}     
\usepackage{enumerate}     
\usepackage{exscale}     
\usepackage{amssymb,amsmath}     
\newcommand{\RR}{\mathbb{R}}     
\newcommand{\NN}{\mathbb{N}}     
\newcommand{\CC}{\mathbb{C}}     
\newcommand{\ZZ}{\mathbb{Z}}     
\newcommand{\EE}{\mathbb{E}}     
\newcommand{\PP}{\mathbb{P}}

\newcommand{\calB}{\mathcal{B}}   
     
\newcommand{\calD}{\mathcal{D}}     
     
\newcommand{\calF}{\mathcal{F}}     
\newcommand{\calG}{\mathcal{G}}     
\newcommand{\calH}{\mathcal{H}}

\newcommand{\calN}{\mathcal{N}}     
\newcommand{\calT}{\mathcal{T}}     
\newcommand{\calX}{\mathcal{X}}

\renewcommand{\r}{\right}     
\renewcommand{\l}{\left}

\newcounter{smalllist}     
     
\newcommand{\vol}{{\mathrm{vol}}}

\newcommand{\tr}{\mathop{\mathrm{tr}}}

\newtheorem{theorem}{Theorem}     
\newtheorem{lemma}{Lemma}[section]     
     
\newtheorem{coro}[lemma]{Corollary}     
\theoremstyle{definition}     
\newtheorem{definition}[lemma]{Definition}     
     
\theoremstyle{remark}     
\newtheorem{remark}[lemma]{Remark}

\begin{document}     

\title[Random Schr\"odinger operators on manifolds]{Random Schr\"odinger   
operators on manifolds}   

\author[D.~Lenz]{Daniel Lenz} \author[N.~Peyerimhoff]{Norbert   
Peyerimhoff} \author[I.~Veseli\'c]{Ivan Veseli\'c}   
\address[D.~Lenz]{Fakult\"at f\"ur Mathematik, TU Chemnitz, 09107   
Chemnitz, Germany} \email{dlenz@mathematik.tu-chemnitz.de}   
\urladdr{www.tu-chemnitz.de/mathematik/analysis/dlenz}   
\address[N.~Peyerimhoff]{Fakult\"at f\"ur   
Mathematik\\Ruhr-Universit\"at Bochum, Germany}   
\email{peyerim@math.ruhr-uni-bochum.de}   
\urladdr{www.ruhr-uni-bochum.de/mathematik10/Norbert.html}   
\address[I.~Veseli\'c]{Postdoctoral research fellow of the Deutsche Forschungsgemeinschaft, {\normalfont visiting the} Department of  Mathematics 253-37, California Institute of Technology, CA 91125, USA} \email{veselic@caltech.edu}   
\urladdr{www.its.caltech.edu/\protect{\symbol{20}}veselic}

\date{\today}     
     
\keywords{integrated density of states, random metrics, random operators,   
Schr\"odinger operators on manifolds, Von Neumann algebra, trace}     
\subjclass[2000]{35J10; 58J35; 82B44}

%%%%%%%%%%%%%%%%%%%%%%%%%%%%%%%%%%%%%%%%%%%%%%%%%%%%%%%%%%%%   
% ABSTRACT   
%%%%%%%%%%%%%%%%%%%%%%%%%%%%%%%%%%%%%%%%%%%%%%%%%%%%%%%%%%%%   
 
\begin{abstract}
We consider a random family of Schr\"odinger operators on a cover $X$
of a compact Riemannian manifold $M = X/\Gamma$. We present several results on
their spectral theory, in particular almost sure constancy of the
spectral components and existence and non-randomness of an integrated
density of states. We also sketch a groupoid based general framework
which allows to treat basic features of random operators in different
contexts in a unified way. Further topics of research are also discussed.
\end{abstract}
  
%\begin{abstract}     
%We consider random Schr\"odinger   
%operator on  manifolds and present several results on their  spectral
%theory. Further topics of research on this subject are also discussed.
%\end{abstract}     

\maketitle

%%%%%%%%%%%%%%%%%%%%%%%%%%%%%%%%%%%%%%%%%%%%%%%%%%%%%%%%%%%%   
% INTRODUCTION   
%%%%%%%%%%%%%%%%%%%%%%%%%%%%%%%%%%%%%%%%%%%%%%%%%%%%%%%%%%%%   
   
\section{Introduction}     \label{Introduction}
     
This paper is devoted to the study of spectral properties of random
operators on manifolds.  Their counterparts in Euclidean geometry play
a prominent part in solid state physics
\cite{Bonch-BruevichEEKMZ-1984,EfrosS-84,Lifschitz-1985,LifshitzGP-88}.
Homogeneous random Hamiltonians are used to describe the propagation
of both quantum mechanical and classical waves in random media.  The
model which most of the mathematical physics literature is devoted to
is a Schr\"odinger operator with random potential, see for instance
\cite{CyconFKS-87, Kirsch-89a,CarmonaL-1990, PasturF-1992,
Stollmann-2001}. Hereby the potentials, more precisely the level of
randomness they describe, may take quite different forms. For
instance, a random potential generated by a Poissonian stochastic
field describes an amorphous medium, whereas a quasi-periodic
potential models a medium whose structure is very close to a
crystalline one. A model where the randomness lies somewhere between
the two just mentioned extremes is the (discrete) Anderson model on $
l^2(\ZZ^d)$ and its continuum counterpart on $ L^2(\RR^d)$, the
alloy-type model.

Characteristic for the last mentioned model is, that while the members
of the random family of operators act on an $L^2$-space over an
continuum configuration space, the homogeneity is expressed via the
action of the discrete group $\ZZ^d$.  We study here analogues of such
operators, which act on more general geometries, namely covering
Riemannian manifolds with compact quotients. Moreover, we allow also
that the randomness enters in the Laplace-Beltrami operator via the
metric.

By the variety of the models which fit under the common roof of random
(Schr\"o\-din\-ger) operators it cannot be expected that they will share
all spectral features in detail. In fact, almost-periodic
Schr\"odinger operators alone illustrate that all measure-theoretic
types of spectrum may occur (see e.g.~Chapter 10 in
\cite{CyconFKS-87}).  However basic spectral features are shared by
all models and can be traced back to rather mild, abstract conditions
on the considered random family of operators. These features include
the non-randomness of the spectrum and existence of a self-averaging
integrated density of states (IDS).

These basic spectral properties are presented in this paper for random
Schr\"odinger operators on manifolds and random Laplace-Beltrami
operators. Furthermore we indicate their origin from the above
mentioned abstract properties using the framework of groupoids and von
Neumann algebras.

Let us say something about the physical picture one has in mind when
studying these models. Random Schr\"odinger operators are used to
study conductance properties of solids. In certain cases it is known
that the boundary plays a important role for the wave transport
properties. This suggest the study of spectral properties of the
Laplace-Beltrami operator of the two-dimensional surface which forms
the boundary of the solid. Since the surface is not perfectly planar,
we want to account for its random holes by using a stochastic field as
the metric.

If one describes wave propagation properties of thin films or layers
--- idealized by surfaces --- it may be necessary to incorporate an
additional random "effective potential", cf.~for instance
\cite{ExnerK-02}.  However this zero order term of the operator is part of the kinetic energy and
is a reminiscence of the fact that the operator on the surface is an
idealization of an operator acting on a very thin layer.

Let us outline the structure of the paper. In the next section we
introduce the random operators on manifolds we are dealing
with. Sections 3 and 4 are devoted to basic spectral properties of
those operators and the existence of an selfaveraging integrated
density of states, while Section 5 and 6 provide insight in the
groupoid/von Neumann algebra background of these results. In Section 7
we specialize to alloy-type models on manifolds. For those we discuss
Wegner estimates. Finally the last section is devoted to some
interesting questions for further study.

%%%%%%%%%%%%%%%%%%%%%%%%%%%%%%%%%%%%%%%%%%%%%%%%%%%%%%%%%%%%   
% OUR MODEL
%%%%%%%%%%%%%%%%%%%%%%%%%%%%%%%%%%%%%%%%%%%%%%%%%%%%%%%%%%%%   

\section{Our model}  \label{Model}

In this section we shortly discuss the geometric situation underlying
our model and then introduce the random Schr\"odinger operators we are
dealing with. The model is a generalization of the model
introduced in \cite{PeyerimhoffV-2002} as has been studied in
\cite{LenzPV-2002?,LenzPV-2002}. Our treatment here mainly follows \cite{LenzPV-2002?,LenzPV-2002}
to which we refer the reader for further details.

\medskip

Our model is based on the following geometric situation: Let $X$ be a
cover of a compact Riemannian manifold $M=X/\Gamma$ where $\Gamma$ is
a discrete, finitely generated subgroup of the isometries of $X$ with
$| \Gamma | = \infty$. Let $g_0$ be the fixed $\Gamma$-periodic smooth
Riemannian metric on the cover $X$, inherited from $M$. Denote by
$\calF$ a precompact connected $\Gamma$-fundamental domain with
piecewise smooth boundary. Furthermore, let $(\Omega, \calB_\Omega,
\PP)$ be a probability space on which $\Gamma$ acts ergodically by
measure preserving transformations $\gamma \colon \Omega \to \Omega,
\gamma \in \Gamma$.  The expectation with respect to $\PP$ is denoted
by $\EE$.
   
\medskip   
   
We will consider two types of random objects over $(\Omega,   
\calB_\Omega, \PP)$. The first is a family of {\em random metrics} on $X$,   
the second is a family of {\em random potentials}. Put together, they will   
give rise to a family of random operators.   
   
\medskip   
   
As for the random metrics, the manifold $X$ is equiped with a family
of metrics $\{g_\omega\}_{\omega\in\Omega}$ with corresponding volume
forms $\vol_\omega$ with the following properties: The map $
(\omega,v) \mapsto g_\omega(v,v)$ is jointly measurable for all
$(\omega,v) \in \Omega \times TX$. There are constants $C_g, C_\rho > 0$ such
that
\begin{equation}      
\label{quasiisom}   
C_g^{-1} g_0(v,v) \le g_\omega(v,v) \le C_g g_0(v,v) \ \forall \,
(\omega,v) \in \Omega \times TX
\end{equation}      
and   
\begin{equation}      
\label{gradbound}   
\vert \nabla_0 \rho_\omega(x) \vert_0 \le C_\rho \ \forall   
\, (\omega,x) \in \Omega \times X.   
\end{equation}      
Here $\nabla_0$ denotes the gradient with respect to $g_0$,
$\rho_\omega$ is the unique smooth density satisfying $d\vol_0 =
\rho_\omega d\vol_\omega$, and $\vert v \vert_0^2 = g_0(v,v)$. The
Ricci curvature of all metrics $g^\omega$ is bounded below by a fixed
constant $K \in \RR$. The metrics are compatible in the sense that the
deck transformations
\[   
\gamma\colon (X,g_\omega) \to (X,g_{\gamma \omega})
\]   
are isometries and that the induced maps 
$$U_{(\omega,\gamma)}\colon L^2(X,\vol_{\gamma^{-1}\omega}) \to
L^2(X,\vol_\omega), \:\;\: (U_{(\omega,\gamma)} f)(x) =
f(\gamma^{-1}x)$$ 
are unitary operators on the family of Hilbert
spaces over the manifolds $\{(X,g_\omega)\}_{\omega\in\Omega}$.
   
\smallskip   
   
As for the random potentials, let $V : \Omega \times X \longrightarrow
[0,\infty)$ be jointly measurable with $V_\omega\equiv V(\omega,\cdot)
\in L^1_{loc}(X,g_\omega)$ for all $\omega \in \Omega$.
Assume furthermore that $V(\gamma \omega, x)= V(\omega,
\gamma^{-1} x)$ for arbitrary $x\in X$, $\omega \in \Omega$ and $\gamma
\in \Gamma$.
   
\medskip   
   
Given a random metric and a random potential, we can now introduce the
corresponding random Schr\"odinger operators as $H_\omega :=
\Delta_\omega + V_\omega$ on   the Hilbert spaces $L^2(X,\vol_\omega)$.
In fact, these unbounded operators are
defined  by means of
quadratic forms. Moreover, they satisfy the \textit{equivariance
condition}
\begin{equation} \label{compcomp}   
H_\omega = U_{(\omega,\gamma)} H_{\gamma^{-1} \omega}   
U_{(\omega,\gamma)}^*,   
\end{equation}   
for all $\gamma \in \Gamma$ and $\omega \in \Omega$. We will refer to
$\{ H_\omega \}$ as a {\em random (Schr\"odinger) operator on the
manifold $X$} and denote this model by {\bf (RSM)}.

%%%%%%%%%%%%%%%%%%%%%%%%%%%%%%%%%%%%%%%%%%%%%%%%%%%%%%%%%   
% GENERAL RESULTS   
%%%%%%%%%%%%%%%%%%%%%%%%%%%%%%%%%%%%%%%%%%%%%%%%%%%%%%%%%   
   
\section{General results}   \label{General}

In this section we discuss measurability properties of our random
operators, almost everywhere constancy of their spectral properties
and we introduce the abstract density of states. Again, the results in
this section are taken from \cite{LenzPV-2002?,LenzPV-2002}.

\medskip

We start by defining the notion of measurability in our setting. 

\begin{definition}     
\label{measurableH}     
Let $D$ be an open subset of $X$. A family of selfadjoint operators
$\{H_\omega\}_\omega$, where the domain of $H_\omega$ is a dense
subspace $\calD_\omega$ of $L^2(D, \vol_\omega)$, is called a {\em
measurable family of operators} if
\begin{equation}  \label{weakmeas}     
\omega\mapsto ( f_\omega, F(H_\omega) f_\omega)_\omega   
\end{equation}   
is measurable  for all measurable $F\colon \RR\rightarrow \CC$ with $|F|$ bounded 
and all $f \colon \Omega\times D\rightarrow \RR$ measurable with
$f_\omega \in L^2(D,\vol_\omega)$, $f_\omega(x) = f(\omega,x)$, for
every $\omega\in \Omega$.  Here, $(\cdot, \cdot)_\omega$ denotes the
inner product on $L^2(D, \vol_\omega)$.
\end{definition}   
   
Our random operators are measurable in this sense as can be seen from
the next theorem.

\begin{theorem}\label{measur}     
A random operator $\{H_\omega\}_{\omega\in\Omega}$ on $X$ as well as
its restriction $\{H_\omega^D\}$ to an arbitrary open set $D \subset
X$ with Dirichlet boundary condition is a measurable family of
operators.
\end{theorem}    

The proof of the theorem is not too complicated but somewhat
technical.  In fact, for technical reasons it is needed that the
$\sigma$-algebra of $\Omega$ is countably generated. In our model this
can be established by changing to an equivalent version of the
defining stochastic processes given by the random potential and the
random metric. This has been done for the potential explicitly in
Remark 2.8 in \cite{LenzPV-2002?}.  Given the measurabilty one can use
the results of \cite{LenzPV-2002?} to establish a result on the
non-randomness of the spectrum.

\begin{theorem} \label{constancy}     
Let $\{H_\omega\}_{\omega\in\Omega}$ be a random operator on $X$. Then   
there exist $\Omega' \subset \Omega$ of full measure and $\Sigma,   
\Sigma_\bullet\subset \RR$, such that $ \sigma(H_\omega)=\Sigma, \quad   
\sigma_\bullet (H_\omega)= \Sigma_\bullet\quad \text{ for all   
}\omega\in \Omega' $ where $\bullet \in \{ disc,ess,ac, sc,pp \}$. Moreover,   
$\Sigma_{disc}=\emptyset$.   
\end{theorem}    

Note that $\sigma_{pp}$ denotes the closure of the set of eigenvalues.

Next, we introduce the {\em (abstract) density of states} for a   
random operator $\{ H_\omega \}$ as the measure on $\RR$, given by   
\begin{equation} \label{absIDS}   
\rho_H(f):= \frac{\EE \l[ \tr \left( \chi_\calF f(H_\bullet) \right)
\r] }{\EE \l[ \vol_\bullet(\calF)\r] }, \quad \text{ for bounded
measurable $f$.}
\end{equation}   
  
The measure $\rho_H$ completely determines the spectral theory of the
direct integral $ H= {\int_\Omega}^\oplus H_\omega $ as can be seen
from the following proposition. Recall that a measure $\phi$ on $\RR$
is a {\em spectral measure} for the selfadjoint operator $H$ with spectral
family $E_H$ if, for Borel measurable $B\subset\RR$, $\phi(B)=0
\Leftrightarrow E_H (B)=0$.

\begin{theorem}\label{AbstIDS}     
The measure $\rho_H$ is a spectral measure for the direct integral
operator \[ H:= {\int_\Omega}^\oplus H_\omega \, d \PP(\omega).\] In
particular, the almost sure spectrum $\Sigma$ coincides with the
topological support $\{ \lambda\in \RR : \rho(\, ]\lambda-\epsilon,
\lambda + \epsilon[\, )>0 \quad \text{for all $\epsilon >0$}\}$ of
$\rho_H$.
\end{theorem}   

\begin{remark}{\rm
Of course, the spectral properties of $H$ will in general be different
and in, fact, ``smoother'' than the spectral properties of the single
operators $H_\omega$. }
\end{remark}

It will turn out that the measure $\rho_H$, its distribution function
$E\mapsto \rho_H(-\infty,E)$ and a certain trace on a suitable von
Neumann algebra are intimately connected. This will be studied in the
next sections.

%%%%%%%%%%%%%%%%%%%%%%%%%%%%%%%%%%%%%%%%%%%%%%%%%%%%%%%%%   
% Shubin-Pastur trace formula   
%%%%%%%%%%%%%%%%%%%%%%%%%%%%%%%%%%%%%%%%%%%%%%%%%%%%%%%%%   

\section{\v Subin-Pastur trace formula}

In this section we discuss how the distribution function of the
abstract density of states can be calculated by an exhaustion
procedure. As the distribution function is essentially given by a trace (see below), this gives effectively a way to calculate a trace by an exhaustion procedure.  This type of formula is associated to the names of Pastur and \v Subin 
after the seminal work \cite{Pastur-1971,Shubin-1979,Shubin-1982}. The material of this section is
taken from \cite{LenzPV-2002}.
   
\medskip

We consider the following normalized eigenvalue counting function for
the restricted random operator $\{ H_\omega^D \}$ on a given open set
$D \subset X$:
\begin{equation} \label{ewcfunc}     
N_\omega^D(\lambda) = \frac{\# \{ i \mid \lambda_i(H_\omega^{D}) < \lambda     
 \}}{\vol_\omega(D) }. \     
\end{equation}     
Obviously, $N_\omega^D$ is a distribution function and has countably
many discontinuity points. Our aim is now to exhaust $X$ by an
increasing set $\{ D^j \}$ of open domains. For well-definedness of
the limit (at all continuity points) we need that the group $\Gamma$
is {\em amenable}. A geometric description of amenability is the existence
of a F{\o}lner sequence. Let us introduce the relevant notions in the
next definition. The function $\phi$ associates to every finite subset
$I \subset \Gamma$ a corresponding open set $\phi(I) = {\rm int}\bigg
( \bigcup_{\gamma \in I} \gamma \overline{\calF} \bigg )$ of $X$, with
the help of the fixed fundamental domain $\calF$ of Section \ref{Model}.
   
\begin{definition} \label{admseq}     
(a) A sequence $\{I_j\}_j$ of finite subsets in $\Gamma$ is called a
  {\em F{\o}lner sequence} if $ \lim_{j \to \infty} \frac{\vert I_j
  \Delta I_j \gamma \vert} {\vert I_j\vert} = 0$ for all $\gamma \in
  \Gamma$.\\ 
(b) A F{\o}lner sequence $\{I_j\}_j$ is called a {\em
  tempered F{\o}lner sequence} if it is monotonously increasing and
  satisfies $\sup_{j \in \NN} \frac{\vert I_{j+1} I_j^{-1}
  \vert}{\vert I_{j+1} \vert} < \infty.$
  \\ 
(c) A sequence $\{D^j\}_j$
  of subsets of $X$ is called {\em admissible} if there exists a
  tempered F{\o}lner sequence $\{I_j\}_j$ in $\Gamma$ with $D^j = \phi
  (I_j)$, $j\in \NN$.
\end{definition}    
   
Lindenstrauss shows in \cite{Lindenstrauss-2001} that every
F{\o}lner sequence has a tempered subsequence, and uses this to prove an
 ergodic theorem for locally compact amenable groups.
 
Using Lindenstrauss results together with suitable heat kernel
estimates, it is shown in \cite{LenzPV-2002?} that, for any admissible
sequence $\{ D^j \}$, and the corresponding normalized eigenvalue
functions $N^j_\omega := N^{D^j}_\omega$, the limit
$$N_H (\lambda)\equiv \lim_{j \to \infty} N^j_\omega(\lambda)$$ 
exists for almost all $\omega \in \Omega$ and for all $\lambda \in
\RR$ with at most countably many exceptions. The limit is again a
distribution function and carries the name \emph{integrated density of
states}.  It agrees with the (nonrandom) distribution function of the
abstract density of states. This result is stated in the following
theorem.
   
\begin{theorem}\label{selfaverIDS}     
Let $\{D^j\}_j$ be an admissible sequence and $\{H_\omega\}_\omega$ be as   
above. There exists a set $\Omega'$ of full measure such that   
\[     
\lim_{j\to\infty} N_\omega^{j}(\lambda) = N_H(\lambda) = \rho_H(\,
]-\infty,\lambda[\, ),
\]     
for every $\omega \in \Omega'$ and every point $\lambda\in \RR$ with   
$\rho_H(\{ \lambda\})=0$.   
\end{theorem}    

Using Dirichlet-Neumann bracketing, we obtain the following corollary as
an immediate consequence of the preceeding theorem (and its proof).

\begin{coro}
For every $\lambda \in \RR$ with $\rho_H(\{ \lambda\})=0$ we have
\[ 
\frac{\EE(\tr(\chi_{]-\infty,\lambda[}(H_\bullet^{\calF,D})))}
{\EE(\vol_\bullet(\calF))} \le N_H(\lambda) \le 
\frac{\EE(\tr(\chi_{]-\infty,\lambda[}(H_\bullet^{\calF,N})))}
{\EE(\vol_\bullet(\calF))},
\]
where $\chi_{]-\infty,\lambda[}(H_\omega^{\calF,\#})$, $\#=D$ or
$\#=N$ denotes the spectral projection of the restricted operator
$H_\omega$ to $\calF$ with Dirichlet, resp., Neumann boundary
condition onto the interval $]-\infty,\lambda[$.
\end{coro}

\section{The density of states as a trace on a von Neumann algebra}
The abstract density of states $\rho_H$ is closely related to a trace
$\tau$ of a suitable von Neumann algebra $\calN$.

In order to describe the von Neumann algebra $\calN$, we first
introduce the concept of a bounded random operator:

\begin{definition} \label{boundrandop}    
A family $\{A_\omega\}_{\omega\in \Omega}$ of bounded operators   
$A_\omega\colon L^2(X,\vol_\omega)\to L^2(X,\vol_\omega)$ is called a   
{\em bounded random operator} if it satisfies:   
\begin{enumerate}[\rm (i)]     
\item $\omega\mapsto \langle g_\omega, A_\omega f_\omega\rangle$ is   
measurable for arbitrary $f,g\in L^2(\Omega\times X, \PP\circ \vol)$.   
\item There exists a $C\geq 0$ with $\|A_\omega\|\leq C$ for almost   
all $\omega \in \Omega$.   
\item For all $\omega\in\Omega, \gamma \in \Gamma$ the equivariance   
condition $ A_\omega = U_{(\omega,\gamma)} A_{\gamma^{-1} \omega}   
U_{(\omega,\gamma)}^*$is satisfied.   
\end{enumerate}    
\end{definition}   
    
Two bounded random operators $\{A_\omega\}_\omega,
\{B_\omega\}_\omega$ are called {\em equivalent},
$\{A_\omega\}_\omega\sim \{B_\omega\}_\omega$, if $A_\omega=B_\omega$
for $\PP$-almost every $\omega\in \Omega$. Each equivalence class of
bounded random operators $\{A_\omega\}_\omega$ gives rise to a {\em
single} bounded operator $A$ on the larger Hilbert space
$L^2(\Omega\times X, \PP\circ \vol_\bullet)$ by $(A f) (\omega,x) :=
A_{\omega} f_{\omega}(x)$, see Appendix A in \cite{LenzPV-2002?}. This allows
us to identify the equivalence class of $\{A_\omega\}_\omega$ with the
bounded operator $A$.
   
Now, the set of bounded random operators form the von Neumann algebra
$\calN$ (see Theorem 3.1 in \cite{LenzPV-2002?}). This von Neumann
algebra possesses a trace \cite{LenzPV-2002?}) which is given by the
following procedure: Choose a measurable $u\colon\Omega\times X \to
\RR^+$ with $\sum_{\gamma \in\Gamma}
u_{\gamma^{-1}\omega}(\gamma^{-1}x) \equiv 1$ on $\Omega\times X$ and
define
\[     
\tau(A) :=   \EE \l[ \tr (u_\bullet A_\bullet) \r]     
\]     
on the set of non-negative operators in $\calN$. This $\tau$ is
independent of $u$ (chosen as above). Under an additional freeness
condition (see Theorem 4.2(b) in \cite{LenzPV-2002?}), $\calN$ is a factor of
type II$_\infty$ and its trace is uniquely determined (up to a constant
multiple).

By \eqref{compcomp} and the last section, the resolvents, spectral
projections and the semigroup associated to $\{H_\omega\}_\omega$ are
all bounded random operators. This can be used to show how $\rho_H$
and $\tau$ are related. Namely, the family
$\{E_\omega(\lambda)\}_\omega$ of spectral projections onto the
interval $]-\infty, \lambda[$ of the random operator
$\{H_\omega\}_\omega$ is an element of $\calN$ and agrees with the
spectral projection of $H:= \int_\Omega^\oplus H_\omega \, d
\PP(\omega) $ onto $]-\infty, \lambda[$.  Hence $\tau(E(\lambda))$ is
well defined.

Choosing $u_\omega(x) = \chi_{\calF} (x)$ we can summarize these
considerations in the following result.

\begin{coro}  Set $\calD\equiv \EE \, (\vol_\bullet \calF)$. 
Then, $\frac{1}{\calD}\, \tau(E(\lambda)) = N_H(\lambda)$.
\end{coro}

%%%%%%%%%%%%%%%%%%%%%%%%%%%%%%%%%%%%%%%%%%%%%%%%%%%%%%%%%   
% The general framework
%%%%%%%%%%%%%%%%%%%%%%%%%%%%%%%%%%%%%%%%%%%%%%%%%%%%%%%%%  

\section{The general framework: Random operators and groupoids}
The abstract considerations in the last section motivate naturally a
{\em more general abstract framework} which covers also random
operators arising in various other settings: random operators on
abelian (or more general) amenable groups (see, e.g.,
\cite{Bellissard-1986,Bellissard-1992,KaminkerX-1987}), random
operators on Bethe lattices (see e.g. \cite{Klein-1996}), on tilings
and Delone sets as (see, e.g.,
\cite{BellissardHZ-2000,Kellendonk-1995,LenzS-2002,LenzS}), on
foliations and on manifolds (see above). This general framework allows
a unified treatment of basic spectral features of random operators in
these different contexts by the use of Connes non-commutative
integration theory \cite{Connes-1979}.  This has been done by the
authors in \cite{LenzPV-2002?}.

There also, as an illustration, the details have been worked out for
random Schr\"odinger operators on manifolds. A corresponding treatment
of quasicrystals has been done in \cite{LenzS-2002,LenzS-2002} (see
\cite{Kellendonk-1995,BellissardHZ-2000} as well).  For almost
periodic operators it can be found in
\cite{Bellissard-1986,Bellissard-1992,KaminkerX-1987,Lenz-1999}.

%A family of random operators $\{ H_\omega \}$ is considered as a
%single operator $H$, acting on a larger Hilbert space $L^2(\Omega
%\times X,\PP \circ \vol_\bullet)$. 

%\medskip

%Common to all models of random operators are the following data:

%\begin{itemize}
%\item a family of operators $(H_\omega)$ and of Hilbert spaces $(\calH_\omega)%$ indexed by $\omega$ in some
%  probability space $(\Omega,\PP)$  such that $H_\omega$ acts on
%  $\calH_\omega$. 
%\item a set  $\Gamma$ of transformations  acting measure preservingly  on $\Omega$ ,

%\item a ``representation`` of $\Gamma$ by unitary operators which
%commutes with $(H_\omega)$.
%\end{itemize}

The abtract framework starts with a measurable groupoid $\calG$ (in
our model $\calG = \Omega \times \Gamma$). The corresponding set of
units, denoted by $\Omega$, carries a measure $\mu$ (which is
invariant w.r.t. $\calG$) and can in our model (RSM) be canonically
identified with the underlying probability space. The groupoid $\calG$
acts on a suitable space $\calX$. This space is a bundle over $\Omega$
with natural projection map $\pi: \calX \to \Omega$ and equiped with a
family $(\alpha^\omega)$ of measures s.t. $\alpha^\omega$ is supported
in $\calX^\omega \equiv \pi^{-1} (\omega)$.

\medskip

Given this setting one can easily define random operators. 
%appropriate measure $\mu \circ \alpha$
%(in our case $\calX = \Omega \times X$ with measure $\PP \circ
%\vol_\bullet$). 
A {\em bounded random operator} is a family of operators $\{ A_\omega
\}$ acting on the Hilbert spaces $L^2(\calX^\omega,\alpha^\omega)$ of
the fibers $\calX^\omega = \pi^{-1}(\omega)$ which satisfies the
following conditions analogous to the ones given in Definition
\ref{boundrandop}
\begin{itemize}

\item the map $\omega \mapsto A_\omega$  is measurable and essentially bounded

\item the family is equivariant, i.e. compatible with the action of $\calG$. 

\end{itemize}

Note that this setting very naturally produces the two main features
of random operators: a family of operators indexed by a probability
space and an equivariance condition.

\medskip

As above, one associates to an equivalence class of a bounded random
operator $\{ A_\omega \}$ in one-one correspondence a single bounded
operator $A$ on the larger Hilbert space $\calH := L^2(\calX,\mu \circ
\alpha)$, and the set of these operators on $\calH$ defines a von
Neumann algebra $\calN$.

\medskip

While this formalism may seem rather abstract and somewhat vague, we
should like to stress that, firstly, it applies to a large number of
models and secondly, the involved quantities can in concrete cases
often easily be identified. We include the following table to make
this transparent. Here, c.m. means counting measure on a discrete
space and L.m. means the Lebesque measure:

\medskip

\begin{tabular}{|l| |c|c|c|c| } \hline     
    Model& Groupoid & Units & Typical fibre & measure 
    \\ & & of $\calG$ &
     $\calX_\omega$ of $\calX$ & $\alpha^\omega$ on $\calX_\omega$ 
     \\
     \hline\hline Anderson & $\Omega \times \ZZ^d $ &
     $\otimes_{\ZZ^d}\RR$ & $\ZZ^d$ & c.m.  
     \\ model & & & & 
     \\ \hline
     Alloy-type & $\Omega \times \ZZ^d$& $ \otimes_{\ZZ^d}\RR $ &
     $\RR^d$ & L.m.  
     \\ model & & & & 
     \\ \hline Lifschitz-Poisson&
     $\Omega \times \RR^d$ &$\otimes_{\RR^d}\RR $ & $\RR^d$& L.m. 
     \\
     model & & & & 
     \\ \hline Random Gaussian & $\Omega \times \RR^d$ &
     $ \otimes_{\RR^d}\RR $ & $\RR^d$ & L.m.  
     \\ potential & & & & 
     \\
     \hline Almost periodic & $\Omega \times \RR^d$ &
      & $\RR^d$ & L.m. 
     \\
     potential & &$\overline{\{f(\cdot-t)| t \in \RR^d\}}$ & &
     \\ \hline (RSM)& $\Omega \times \Gamma$ & $
     \Omega$ & $(X,g_\omega)$ & $\vol_\omega$ 
     \\ \hline Delone system
     & $\Omega \times \RR^d$ & $\Omega$ & $\omega$ &c.m.  
     \\ \hline
     Tiling $\calT$ & $\calG(\calT)$ & Tiling space & Tiling & c.m.
     \\ \hline
     Foliation& $\calG (X,\calF)$ & $X$ & $\calF_\omega$ &
     $\vol_{\calF_\omega}$ \\ \hline
%& Discrete magnetic & $\{pt\} \times \Gamma$ & $\{pt\}$ & $G$ graph &
%c.m.  \\   
%    &  Laplacian      &                        &       &           &        %           \\ \hline  
%12.& Random necklace& $\Omega\times\ZZ$ &  & $X_\omega$ & $\vol_\omega$  \\ \hline  
\end{tabular}  

\medskip

As can be seen in the diagram, in many models the groupoid is given by
the product of a group with the underlying probability space.

\smallskip

Now, given this framework as well as some additional technical
conditions (see Definition 2.6, Theorem 4.2(b) and Lemma 5.6 in
\cite{LenzPV-2002?}) the following properties of the von Neumann
algebra $\calN$ and an affiliated random family $\{ H_\omega \}$ can be
established \cite{LenzPV-2002?}:

\begin{enumerate}[(i)]

\item Almost sure constancy of the spectral components of $\{ H_\omega \}$, 

\item  absence of discrete spectrum for almost every $\{ H_\omega \}$, 

\item existence of a faithful  trace  $\tau$ on $\calN$, 
which gives rise to a spectral measure of the family $\{ H_\omega \}$.

\end{enumerate}

\medskip

%To derive properties listed under (i) and (ii) above, one has to
%require the following condition or show that they are immanent in the
%model.
 
% \begin{enumerate}[(a)]
%\item 
%Firstly, a strong form of separability of the probability space is
%necessary. More precisely, the $\sigma$-algebras $\calB_{\calX}$ and
%$\calB_{\Omega}$ have to be generated by a countable family of sets,
%5all of which have finite measure.
%\item 
%There exists a measurable strictly positive function $u$ on $\calX$  such that% the integral $u$ over the $\calG$-orbit of any $p\in \calX$ equals 1. \textbf{%Ist das mit dem Orbit richtig formuliert?}
%\item
%The groupoid satisfies a certain freeness condition. For instance, it
%is sufficient if for any unit $\omega \in \Omega$, the intersection of
%the preimages of $\omega$ under the range and source functions
%contains only $\omega$ itself.
%\item
%The groupoid has to have "infinite volume" in the following sense:  There exis%ts a sequence $(f_n)$ of measurable, non-negative functions on $\calG$ with 
%\begin{align} 
%\calG=\bigcup\nolimits_n \{g: f_n(g)>0\},\|f_n\|_{\infty}\rightarrow 0
%\text{ as } n\rightarrow \infty \text{ and } \nu(\tilde{f_n}) \equiv 1
%\text{ for all } n \in \NN.  \nonumber
%\end{align}
%Here $\tilde{f}(g)=f(g^{-1})$.
%\end{enumerate}

Under suitable additional conditions one can even provide a trace
formula, or more precisely:
\begin{enumerate}
\item[(iv)] 
An explicit way to calculate the trace using an exhaustion procedure.
\end{enumerate}

To infer such a formula, one has to have information not only on the
groupoid $\calG$, but also on the space $\calX$ and on the family
of operators $(H_\omega)$ on the Hilbert space $L^2(\calX_\omega,
m_\omega)$.  As for the properties of $\calX$, one has to be able to
control boundary effects. More precisely, a exhausting sequence of
subsets of $\calX$ must exist, such that in the limit the "area" of
the boundaries becomes negligible when compared to the volume of the
subsets in the exhaustion sequence.  Furthermore, an appropriate
ergodic theorem (for $\calG$) must be at disposal. It ensures that
spectral quantities are averaged out as one proceeds along the
exhaustion sequence.  The operators $H_\omega$ acting on
$L^2(\calX_\omega, m_\omega)$ --- more precisely, certain auxiliary
functions of them --- must satisfy a "finite range" condition or a
"principle of not feeling the boundary".  The discrete Laplacians on
graphs and continuous Laplacians are examples of such operators.

%We think that this abstract framework is general enough to cover also,
%under certain additional conditions, other cases, for example, almost
%periodic Euclidean Schr\"odinger operators, discrete magnetic
%Laplacians, foliations or random necklaces (was ist das eigentlich?).

\section{Wegner estimates and  continuity properties of the IDS }
\label{Continuity}
In this section we discuss Wegner estimates and continuity properties
of the integrated density of states in our model. The material is
taken from \cite{LenzPPV}.

\medskip

Let $\{ H_\omega \}$ be our random operator. Let $J\subset \Gamma$ be
a finite subset of $\Gamma$, $D:= \phi(J)$ and denote by $P_\omega^{J}
$ the spectral family of the restricted operator $H_\omega^D$ with
Dirichlet boundary condition. An estimate of the form
\begin{equation}\label{WE}
\EE (\tr (P_\bullet^J ([E-\epsilon,E+\epsilon]))) \leq C \epsilon^\alpha 
|J|^\beta
\end{equation}
with $C,\alpha,\beta>0$ and $E$ in a suitable interval $I$ in $\RR$ is
called a \textit{Wegner estimate} after the work \cite{Wegner-81} of
Wegner. Wegner estimates play a crucial role in proofs of
localization 
(see e.g.~\cite{FroehlichS-83,DreifusK-89,CombesH-94b,Stollmann-2001,GerminetK-2001b}). 
Moreover,  they can be used to establish
continuity properties of the integrated density of states. For
example, if \eqref{WE} holds with $\beta=1$ and $C$ and $\alpha$
independent of $J$ for all energies in a certain energy interval $I$,
then the results of the last section, in particular, Theorem
\ref{selfaverIDS} imply
\begin{equation}\label{LS}
N([E-\epsilon,E+ \epsilon]) \leq C \epsilon^\alpha
\end{equation}
for all $E\in I$. This means that $N$ is H\"older continuous with
exponent $\alpha$.
\medskip

To obtain Wegner estimates in our setting, we must further specialize
our model. In the following we present two specific examples where it
is possible to derive Wegner estimates.  In both models one part of
the energy, either the kinetic or the potential, is of alloy type. In
the fist one the potential has this structure, while in the second it
is the metric of the Laplace-Beltrami operator.

We call a random operator $\{ H_\omega \}$ alloy potential model if
the potential $V_\omega$ is relatively small with respect to
$\Delta_\omega$, uniformly in $\omega$ and can be decomposed into two
parts $V_\omega=V_{per}+V^\omega$. The first one $V_{per}$ is
$\Gamma$-periodic, while the second one is the generalization of the
well known Euclidean alloy-type potential 
\begin{eqnarray*} V^\omega(x)=
\sum_{\gamma\in\Gamma} q_\gamma(\omega) v(\gamma^{-1}x).  
\end{eqnarray*} 
Here
the \emph{coupling constants} $q_\gamma, \gamma\in\Gamma$ are a
collection of independent, identically distributed random
variables. The distribution measure $\tilde\mu$ of $q_e$ (which
coincides with the distribution measure of $q_\gamma$ for every
$\gamma$) is assumed to have a density $f \in L_c^\infty$ with respect
to the Lebesgue measure.  The \emph{single site potential} $v\in
L_c^\infty$ is a function satisfying $v \ge \tilde \kappa \chi_\calF$
for some $\tilde \kappa >0$.
  
In this model we assume the metric to be non-random, i.e.~let
   $g_\omega = g_0$ for all $\omega \in \Omega$.  
\smallskip

The other model where a Wegner estimate can be derived is a random
operator $\{ H_\omega \}$ with alloy-type metric. Here we again have
to assume that the potential $V_\omega$ is a relatively small
perturbation with respect to $\Delta_\omega$, uniformly in $\omega$.

The metric is random and resembles the alloy-type potential
above. Namely, for a collection of independently identically
distributed random variables $r_\gamma\colon \Omega\to \RR, \gamma \in
\Gamma$, whose distribution measure $\mu$ has a compactly supported
density $f$ in the Sobolev space $W_1^1(\RR)$ and a \emph{single site
deformation} $u\in C_c^\infty(X)$ with $u\geq \kappa \chi_\calF,
\kappa >0$ define for each $\omega \in \Omega$
$$   
g_\omega(x)= \bigg(\sum_{\gamma\in \Gamma}\, e^{r_\gamma(\omega)} \,   
u(\gamma^{-1} x) \bigg) g_0(x).    
$$   
In this model we assume that the stochastic process $V_\omega$ is independent of $\{r_\gamma\}_\gamma$.

\medskip

In both cases, a Wegner estimate can be established in a suitable
energy region. We refer the reader to \cite{LenzPPV} for further
details.

%%%%%%%%%%%%%%%%%%%%%%%%%%%%%%%%%%%%%%%%%%%%%%%%%%%%%   
% Groupoids and general random operators   
%%%%%%%%%%%%%%%%%%%%%%%%%%%%%%%%%%%%%%%%%%%%%%%%%%%%%   

%%%%%%%%%%%%%%%%%%%%%%%%%%%%%%%%%%%%%%%%%%%%%%%%%%   
% Ausblick und Vorstellung angepeilter Problemstellungen   
%%%%%%%%%%%%%%%%%%%%%%%%%%%%%%%%%%%%%%%%%%%%%%%%%%   
   
\section{Outlook}   
   
The line of research started in \cite{LenzPV-2002?,LenzPV-2002} and (partly)
summarized above can be pursued in various directions. Two of the main
directions are

\begin{itemize}
\item[($\alpha$)] the study of more specific spectral
features for more specialized models on manifolds,

\smallskip
 
\item[($\beta$)] to extend the investigation to other models and geometries. 
\end{itemize}

In both of these directions, various questions are immediate. We
summarize some of them shortly in the remainder of the section.

\medskip

\noindent
As for ($\alpha$), the following topics deserve special mentioning:

\begin{itemize}   
\item[$(*)$] Lifshitz tails for our models,

\smallskip

\item[$(*)$] localization phenomena of Laplacians on manifolds with  
randomly perturbed metrics,   

\smallskip

\item[$(*)$] decay properties of eigenfunctions,

\smallskip

\item[$(*)$] random Schr\"odinger operators with magnetic fields.

\end{itemize}   
   
\medskip

\noindent
As for ($\beta$), the following directions are of particular
interest to us:

\begin{itemize}   
\item[$(*)$] Integrated density of states for foliations: existence
and a \v{S}ubin trace formula,

\smallskip

\item[$(*)$] a \v{S}ubin trace formula for random graphs,    

\smallskip

\item[$(*)$] integrated density of states for divergence type operators on manifolds.  

\end{itemize}   
   
Some of these topics are subject of our current research.

\newcommand{\etalchar}[1]{$^{#1}$}
\def\cprime{$'$} \def\cprime{$'$} \def\cprime{$'$}


\begin{thebibliography}{BBEE{\etalchar{+}}84}

\bibitem[BBEE{\etalchar{+}}84]{Bonch-BruevichEEKMZ-1984}
V.L. Bonch-Bruevich, R.~Enderlein, B.~Esser, R.~Keiper, A.G. Mirnov, and I.P.
  Zyvagin.
\newblock {\em Elektronentheorie ungeordneter Halbleiter}.
\newblock VEB Deutscher Verlag der Wissenschaften, 1984.
\newblock Russisches Original: Moskau, Nauka,1981.

\bibitem[Bel86]{Bellissard-1986}
J.~Bellissard.
\newblock ${K}$-theory of ${C}\sp \ast$-algebras in solid state physics.
\newblock In {\em Statistical mechanics and field theory\thinspace:
  mathematical aspects (Groningen, 1985)}, pages 99--156. Springer, Berlin,
  1986.

\bibitem[Bel92]{Bellissard-1992}
J.~Bellissard.
\newblock Gap labelling theorems for {S}chr\"odinger operators.
\newblock In {\em From number theory to physics (Les Houches, 1989)}, pages
  538--630. Springer, Berlin, 1992.

\bibitem[BHZ00]{BellissardHZ-2000}
J.~Bellissard, D.~J.~L. Herrmann, and M.~Zarrouati.
\newblock Hulls of aperiodic solids and gap labeling theorems.
\newblock In {\em Directions in mathematical quasicrystals}, volume~13 of {\em
  CRM Monogr. Ser.}, pages 207--258. Amer. Math. Soc., Providence, RI, 2000.

\bibitem[CFKS87]{CyconFKS-87}
H.~L. Cycon, R.~G. Froese, W.~Kirsch, and B.~Simon.
\newblock {\em {Schr\"odinger} Operators with Application to Quantum Mechanics
  and Global Geometry}.
\newblock Text and Monographs in Physics. Springer, Berlin, 1987.

\bibitem[CH94]{CombesH-94b}
J.-M. Combes and P.D. Hislop.
\newblock Localization for some continuous, random {Hamiltionians} in
  d-dimensions.
\newblock {\em J. Funct. Anal.}, 124:149--180, 1994.

\bibitem[CL90]{CarmonaL-1990}
R.~Carmona and J.~Lacroix.
\newblock {\em Spectral Theory of Random {Schr\"odinger} Operators}.
\newblock Birkh\"auser, Boston, 1990.

\bibitem[Con79]{Connes-1979}
A.~Connes.
\newblock Sur la th\'eorie non commutative de l'int\'egration.
\newblock In {\em Alg\`ebres d'op\'erateurs (S\'em., Les Plans-sur-Bex, 1978)},
  pages 19--143. Springer, Berlin, 1979.

\bibitem[EK]{ExnerK-02}
P.~Exner and S.~Kondej.
\newblock Bound states due to a strong $\delta$ interaction supported by a
  curved surface.
\newblock submitted to J. Phys. {\bf A}, math-ph/0207025 at arXiv.org.

\bibitem[ES84]{EfrosS-84}
A.~L. Efros and B.~I. Shklovski.
\newblock {\em Electronic Properties of Doped Semi-conductors}.
\newblock Springer, Berlin, 1984.

\bibitem[FS83]{FroehlichS-83}
J.~Fr\"ohlich and T.~Spencer.
\newblock Absence of diffusion in the {Anderson} tight binding model for large
  disorder or low energy.
\newblock {\em Commun. Math. Phys.}, 88:151--184, 1983.

\bibitem[GK01]{GerminetK-2001b}
F.~Germinet and A.~Klein.
\newblock A characterization of the {Anderson} metal-insulator transport
  transition.
\newblock Preprint, www.ma.utexas.edu/mp\_arc/, 2001.

\bibitem[Kel95]{Kellendonk-1995}
J.~Kellendonk.
\newblock Noncommutative geometry of tilings and gap labelling.
\newblock {\em Rev. Math. Phys.}, 7(7):1133--1180, 1995.

\bibitem[Kir89]{Kirsch-89a}
W.~Kirsch.
\newblock Random {Schr\"odinger} operators.
\newblock In H.~Holden and A.~Jensen, editors, {\em {Schr\"odinger} Operators},
  Lecture Notes in Physics, {\bf 345}, Berlin, 1989. Springer.

\bibitem[Kle96]{Klein-1996}
A.~Klein.
\newblock Spreading of wave packets in the {A}nderson model on the {B}ethe
  lattice.
\newblock {\em Comm. Math. Phys.}, 177(3):755--773, 1996.

\bibitem[KX87]{KaminkerX-1987}
J.~Kaminker and J.~Xia.
\newblock The spectrum of operators elliptic along the orbits of ${\bf {r}}\sp
  n$ actions.
\newblock {\em Comm. Math. Phys.}, 110(3):427--438, 1987.

\bibitem[Len99]{Lenz-1999}
D.~H. Lenz.
\newblock Random operators and crossed products.
\newblock {\em Math. Phys. Anal. Geom.}, 2(2):197--220, 1999.

\bibitem[LGP88]{LifshitzGP-88}
I.~M. Lifshitz, S.~A. Gredeskul, and L.~A. Pastur.
\newblock {\em Introduction to the Theory of Disordered Systems}.
\newblock Wiley, New York, 1988.
\newblock Russian original: Nauka, Moscow, 1982.

\bibitem[{Lif}85]{Lifschitz-1985}
Lifschitz {Memorial} {Issue}.
\newblock {\em J. Statist. Phys.}, 38(1-2), 1985.

\bibitem[Lin01]{Lindenstrauss-2001}
E.~Lindenstrauss.
\newblock Pointwise theorems for amenable groups.
\newblock {\em Invent. Math.}, 146(2):259--295, 2001.

\bibitem[LPPV]{LenzPPV}
D.~Lenz, N.~Peyerimhoff, O.~Post, and I.~Veseli\'c.
\newblock Continuity properties of the integrated density of states on
  manifolds.
\newblock in preparation.

\bibitem[LPVa]{LenzPV-2002?}
D.~Lenz, N.~Peyerimhoff, and I.~Veseli\'c.
\newblock Integrated density of states random metrics on manifolds.
\newblock submitted, September 2002.

\bibitem[LPVb]{LenzPV-2002}
D.~Lenz, N.~Peyerimhoff, and I.~Veseli\'c.
\newblock Von {Neumann} algebras, groupoids and the integrated density of
  states.
\newblock (math-ph/0203026 on arXiv.org), submitted, March 2002.

\bibitem[LS]{LenzS}
D.~H. Lenz and P.~Stollmann.
\newblock An ergodic theorem for delone dynamical systems and existence of the
  density of states.
\newblock in preparation.

\bibitem[LS02]{LenzS-2002}
Daniel Lenz and Peter Stollmann.
\newblock Quasicrystals, aperiodic order, and groupoid von {N}eumann algebras.
\newblock {\em C. R. Math. Acad. Sci. Paris}, 334(12):1131--1136, 2002.

\bibitem[Pas71]{Pastur-1971}
L.~A. Pastur.
\newblock Selfaverageability of the number of states of the {S}chr\"odinger
  equation with a random potential.
\newblock {\em Mat. Fiz. i Funkcional. Anal.}, (Vyp. 2):111--116, 238, 1971.

\bibitem[PF92]{PasturF-1992}
L.~A. Pastur and A.~L. Figotin.
\newblock {\em Spectra of Random and Almost-Periodic Operators}.
\newblock Springer Verlag, Berlin, 1992.

\bibitem[PV02]{PeyerimhoffV-2002}
N.~Peyerimhoff and I.~Veseli\'c.
\newblock Integrated density of states for ergodic random {Schr\"{o}dinger}
  operators on manifolds.
\newblock {\em Geom. Dedicata}, 91(1):117--135, 2002.

\bibitem[Sto01]{Stollmann-2001}
P.~Stollmann.
\newblock {\em Caught by disorder: A Course on Bound States in Random Media},
  volume~20 of {\em Progress in Mathematical Physics}.
\newblock Birkh\"auser, 2001.

\bibitem[{\v{S}}ub79]{Shubin-1979}
M.~A. {\v{S}}ubin.
\newblock Spectral theory and the index of elliptic operators with
  almost-periodic coefficients.
\newblock {\em Uspekhi Mat. Nauk}, 34(2(206)):95--135, 1979.

\bibitem[{\v{S}}ub82]{Shubin-1982}
M.~A. {\v{S}}ubin.
\newblock Density of states of self adjoint operators with almost periodic
  coefficients.
\newblock {\em Amer. Math. Soc. Translations}, 118:307--339, 1982.

\bibitem[vDK89]{DreifusK-89}
H.~von Dreifus and A.~Klein.
\newblock A new proof of localization in the {Anderson} tight binding model.
\newblock {\em Commun. Math. Phys.}, 124:285--299, 1989.

\bibitem[Weg81]{Wegner-81}
F.~Wegner.
\newblock Bounds on the {DOS} in disordered systems.
\newblock {\em Z. Phys. B}, 44:9--15, 1981.

\end{thebibliography}
\end{document}